\documentclass[%
 aip,
 amsmath,amssymb,
 reprint,%
]{revtex4-1}

\usepackage{graphicx}% Include figure files
\usepackage{dcolumn}% Align table columns on decimal point
\usepackage{bm}% bold math
\usepackage[utf8]{inputenc}
\usepackage[T1]{fontenc}
\usepackage{mathptmx}
\usepackage{etoolbox}

\makeatletter
\def\@email#1#2{%
 \endgroup
 \patchcmd{\titleblock@produce}
  {\frontmatter@RRAPformat}
  {\frontmatter@RRAPformat{\produce@RRAP{*#1\href{mailto:#2}{#2}}}\frontmatter@RRAPformat}
  {}{}
}%
\makeatother

\usepackage{xcolor}
\newcommand{\bb}{\color{blue}}
\newcommand{\re}{\color{red}}

\newcommand{\dgr}{\color{darkgreen}}

\definecolor{darkgreen}{rgb}{0.0, 0.5, 0.0}
\definecolor{applegreen}{rgb}{0.55, 0.71, 0.0}
\definecolor{bostonuniversityred}{rgb}{0.8, 0.0, 0.0}
\definecolor{capri}{rgb}{0.0, 0.75, 1.0}
\definecolor{cyan}{rgb}{0.0, 1.0, 1.0}
\definecolor{darkred}{rgb}{0.55, 0.0, 0.0}
\definecolor{electricultramarine}{rgb}{0.25, 0.0, 1.0}
\definecolor{green(colorwheel)(x11green)}{rgb}{0.0, 1.0, 0.0}
\definecolor{malachite}{rgb}{0.04, 0.85, 0.32}
\definecolor{ao}{rgb}{0.0, 0.0, 1.0}

\usepackage[normalem]{ulem} % strike text through

\usepackage{soul} % per strikethrough
\usepackage{lineno}
\usepackage[section]{placeins} % per forzare posizione delle sezioni
\begin{document}

\preprint{AIP/123-QED}

\title{The deterministic excitation paradigm and 
		the late Pleistocene glacial terminations}
% Force line breaks with \\
%\author{A. Author}
% \altaffiliation[Also at ]{Physics Department, XYZ University.}%Lines break automatically or can be forced with \\
%\author{B. Author}%
% \email{Second.Author@institution.edu.}
%\affiliation{ 
%Authors' institution and/or address%\\This line break forced with \textbackslash\textbackslash
%}%

\author{Stefano Pierini}
% \homepage{http://www.Second.institution.edu/~Charlie.Author.}
\affiliation{%
Department of Science and Technology, Parthenope University of Naples, Italy (stefano.pierini@uniparthenope.it)%\\This line break forced% with \\
}%

%\date{\today}% It is always \today, today,
             %  but any date may be explicitly specified

\begin{abstract}
A deterministic excitation (DE) paradigm is formulated, according to which the late Pleistocene glacial terminations correspond to the excitation, by the orbital forcing, of nonlinear relaxation oscillations (ROs) internal to the climate system in the absence of any stochastic parameterization. Specific threshold crossing rules  parameterizing the activation of internal climate feedbacks leading to RO excitations are derived according to the DE assumption. They are then applied to an energy balance model describing the fluctuations induced by realistic orbital forcing on the glacial state. The timing of the glacial terminations thus obtained in a reference simulation is found to be in good agreement with proxy records. A sensitivity analysis insures the robustness of the timing. The potential irrelevance of noise allowing DE to hold is discussed, and a possible explanation of the 100-kyr cycle problem based on DE is outlined. In conclusion, the DE paradigm characterizes in one of the simplest possible ways the link between orbital forcing and glacial terminations implied by the Milankovitch hypothesis.
\end{abstract}

\maketitle

\begin{quotation}
Oscillations of the climate system lasting about 100 kyr have been revealed by proxy data in the late Pleistocene. They evidence strong changes in the global ice volume, CO$_2$ concentration, surface temperature, etc., and are all composed of a long glacial state, an abrupt shift to an interglacial state (like the Holocene in which we are living) and a slow return to a new glacial state. Milutin Milankovitch hypothesized a century ago that the glacial-interglacial transitions were paced by an increased solar radiation received in the summer in the northern hemisphere due to the orbital forcing. This is fascinating, as it points to a potentially predictable phenomenon in a highly chaotic system such as climate. In this work, a dynamical paradigm denoted deterministic excitation is formulated, and successfully tested, with the aim of characterizing in the simplest possible way the link between orbital forcing and glacial-interglacial transitions implied by the Milankovitch hypothesis.

\iffalse
{\bb In addition, each article in Chaos is preceded by a lead paragraph targeted at non-specialist readers. This paragraph provides a sense of the context of the work and conveys the primary results in language that is accessible to the journal's broad interdisciplinary audience.

The ``lead paragraph'' is encapsulated with the \LaTeX\ 
\verb+quotation+ environment and is formatted as a single paragraph before the first section heading. 
(The \verb+quotation+ environment reverts to its usual meaning after the first sectioning command.) 
Note that numbered references are allowed in the lead paragraph.
%
The lead paragraph will only be found in an article being prepared for the journal \textit{Chaos}
\fi
%}
\end{quotation}

%====================================================
%====================================================
%====================================================
\section{\label{sec:intro}Introduction\protect\\ 
%The line break was forced \lowercase{via} \textbackslash\textbackslash
} %\cite{ditl_ea20,bosio_ea22,tzip_ea22,lisiecki05,hansen_ea10,hansen_ea13}
\begin{figure}[ht]
\centering
\includegraphics[width=.45\textwidth,scale=.7]{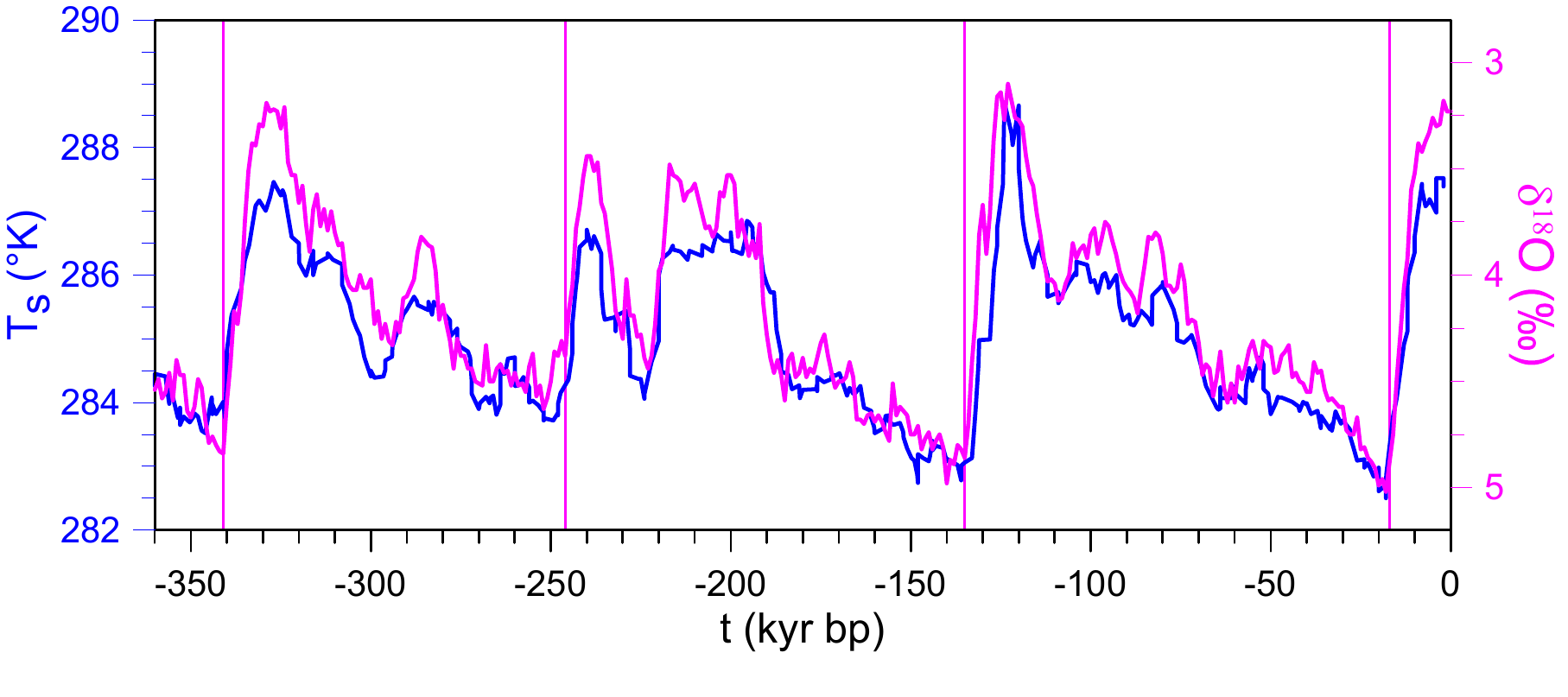}
\caption{Magenta line: LR04 benthic $\delta^{18}$O stack constructed by the graphic correlation of 57 globally distributed benthic $\delta^{18}$O records (from \citet{lisiecki05}; the magenta vertical lines mark the onset of the abrupt glacial-interglatial transitions). Blue line: global surface temperature estimate (from \citet{hansen_ea13}).}
\label{fig1}
\end{figure}

In the late Pleistocene (LP) ice age, glacial periods lasting about 100 kyr terminate abruptly \cite{konij_ea15} and are followed by much shorter interglacials \cite{berger_ea16} which, in turn, relax more slowly to a new glacial period. Fig.~\ref{fig1} shows the typical saw-tooth shape resulting, for example, in a benthic $\delta^{18}$O record (\citet{lisiecki05}) and in a global surface temperature estimate (\citet{hansen_ea13}). The timing of these cycles is believed to be controlled by the orbital forcing according to the \citet{milankovitch20, milankovitch41} hypothesis \cite[e.g.,][]{hays_ea76,GhCh87,bradley99, ruddiman14}. The extreme complexity of the phenomenon has promoted the development of numerous low-order conceptual glacial-cycle models which incorporate the concepts and methods of the theory of autonomous, nonautonomous and random dynamical systems (\citet{Ghil.1994,ghil19}, Boers \textit{et al.}\cite{boers_ea22}). 

These models include 1-dimensional energy balance models (e.g., Refs.\cite{sellers69,budyko69,held74,north75,ghil76,ghil84}) and more complex low-order models that exhibit multistability and/or limit cycles --sometimes in the form of relaxation oscillations (ROs)-- that mimic the glacial-interglacial variability. \citet{riechers_ea22} provide a review of conceptual models with only 1-3 variables (see in particular Table A1 therein). \citet{crucifix_12} reviews  conceptual models in Pleistocene climate theories based on the concepts of relaxation oscillator and excitability. \citet{alexandrov_ea21} provide a review of the same class of models, with particular reference to the effects of stochastic parameterizations. 

One of the main features of the Milankovitch hypothesis is that the abrupt glacial terminations occur in synchrony with an increased solar radiation received in the summer in the northern hemisphere due to the orbital forcing. However, these changes are by far too weak to account for the considerably higher mean surface temperatures typical of the interglacials. Thus, any conceptual glacial-cycle model must include a suitable parameterization of the vigorous nonlinear positive feedbacks internal to the climate system (e.g., the atmospheric CO$_2$, ice-albedo, sea ice feedbacks, etc.) that, once triggered by the increased insolation, can lead to the abrupt reduction of ice sheet mass and increase in surface mean temperature evidenced by the proxies. Moreover, the models may also include a stochastic parameterization (noise) of the effects of internal climate variability on time scales that are smaller than those of the glacial-interglacial variability \cite{hasselmann76,imkvst01,imkmon02,wilks10}. 

In this regard, several deterministic conceptual models subjected to the orbital forcing in which specific threshold crossing rules are prescribed (e.g., Refs.\cite{calder74,imbrie_ea93,paillard98,paillard01,tziperman-gildor03,huybers07,dietlevsen09,imbrie_ea11,parrenin-paillard03,parrenin-paillard12,tzedakis_ea17,berends_ea21,leloup-paillard22}%in which noise --if present-- plays a secondary role
) lead to a correct timing of the glacial-interglacial transitions. For example, in \citet{paillard98} a system possesses a full glacial, a mild glacial and an interglacial state; moreover, rules parameterizing nonlinear internal feedback mechanisms are prescribed in such a way that the transition from one state to the other occurs if the insolation, or the global ice volume, exceed or fall below some given thresholds. Despite the simplicity of the model, the timing of the glacial terminations obtained by applying specific threshold crossing rules to the real (normalized) insolation --without any stochastic component-- is in good agreement with proxy data. 

Although climate is an extremely complex and chaotic system, it is fascinating and somewhat surprising that deterministic conceptual models including simple threshold crossing rules can account for the timing of the abrupt glacial-interglacial transitions, in so validating the original Milankovitch hypothesis. 

Thus, exploring further such rules is important to elucidate and characterize the link between the orbital forcing and the timing of the glacial terminations, and subsequent glacial-interglacial transitions. It is worth noting that those rules take different forms depending on the conceptual model and, despite the simplicity of the latter, they are often overshadowed by the technical details of the model study. It would therefore be desirable to obtain a characterization of the threshold crossing rules parameterizing the deglaciations in one of the simplest possible dynamic scenarios.

In this context, in the present study the deterministic excitation (DE) paradigm based on the concept of RO and excitability is formulated, and is tested for the last four glacial terminations of the LP with an energy balance model describing the fluctuations about the glacial state induced by realistic orbital forcing. The aim of the study is to characterize in the simplest possible way the link between orbital forcing and glacial terminations through specific threshold crossing rules. 

Such rules are very idealized (possible improvements are suggested) and the model used is the most idealized one can conceive. But it is thanks to the very simplicity of the approach and to the good agreement found between the obtained glacial termination timing and that derived from proxy records, that the results of this study may be considered as a useful minimal conceptual tool for the interpretation of the glacial terminations. Less idealized approaches (e.g., including also rules parameterizing the effect of orbital changes on glacial inceptions, noise, etc.) will provide a more realistic --but perhaps less intuitive-- description of the phenomenon.

The paper is organized as follows. In Sect.~\ref{sec:dep} the DE paradigm is formulated and the corresponding threshold crossing rules are defined. In Sect.~\ref{sec:res} the results are presented: in Sect.~\ref{sec:rs} the use of an energy balance model is motivated, the orbital forcing is discussed and a reference simulation is presented, while in Sect.~\ref{sec:se} sensitivity experiments are discussed. In Sect.~\ref{sec:disc} the role of noise in the glacial-interglacial transitions (Sect.~\ref{sec:noise}) and the 100-kyr cycle problem (Sect.~\ref{sec:cp}) are discussed with reference to the DE paradigm. In Sect.~\ref{sec:conc} a summary is presented and conclusions are drawn. Finally, in Appendix A the derivation of the model and a bifurcation analysis of the corresponding autonomous system are given.
%\vspace{.7cm}

%====================================================
%====================================================
%====================================================
\section{\label{sec:dep}The deterministic excitation paradigm}
Here the concepts of RO and excitability are first recalled. The DE paradigm and the corresponding threshold crossing rules to be applied in Sect.~\ref{sec:res} will then be defined.

In a dissipative nonlinear dynamical system a RO is a large-amplitude oscillation that connects a basic state to an unstable excited state, which is then followed by a spontaneous, slow return to the original state (the basic state can be either an equilibrium point, a small amplitude limit cycle or even a chaotic attractor with limited extension in phase space). 

The ROs are self-sustained in a given parameter range of the autonomous system, otherwise they can be excited by a suitable external time-dependent forcing, whether deterministic or random. The basic idea is that the ROs are structured in phase space even in the parameter range in which they are not self-sustained, in which case a suitable external forcing can let them arise. In the range in which the ROs are self-sustained  the system is said to be a \textit{relaxation oscillator} while in the complementary range one has an \textit{excitable system } (e.g., \citet{crucifix_12}). The transformation from a relaxation oscillator to an excitable system, and viceversa, can be obtained by a mere change in parameter.

As far as the nature of the external forcing is concerned, in general if the forcing exciting the ROs is random one speaks of coherence resonance (CR, e.g., \citet{pikovsky97}). Here the DE paradigm refers, instead, to the case in which ROs are excited by a deterministic time-dependent forcing. In the present paleoclimate context understanding whether the glacial-interglacial transitions are excited by the deterministic orbital forcing or, rather, by a predominantly random forcing is of fundamental importance (see Sect.~\ref{sec:noise}). But on the other hand, CR and DE are different manifestations of the same excitation mechanism.

Here it is assumed that the oscillations found in paleorecords describing the glacial-interglacial transitions, and viceversa (Fig.~\ref{fig1}), can be interpreted as ROs emerging through the DE mechanism in an excitable system. The validity of this hypothesis will be verified in Sect.~\ref{sec:res}.

To illustrate the concept of RO and its excitability and  to motivate the rules that parameterize the excitation, an example provided by the oceanic excitable low-order quasigeostrophic model of \citet{pieriniJPO2011} will now be presented. Such model was developed to analyze on a conceptual level an oceanic problem (the Kuroshio Extension interannual-to-decadal intrinsic variability) that was previously investigated in much more realistic model studies (e.g.,\cite{pieriniJPO2006,pieriniDRJPO2009,pieriniJC2014}). The aim was, therefore, to study a phenomenon that is completely different from the one investigated here; nevertheless, that model can be considered as a useful generic dynamical tool to illustrate typical features of excitable systems\cite{pieriniJPO2014,PGC16,PCG18,pg21} and in this perspective it will now be used.

The system is composed of four nonlinear coupled ODEs for the components of the vector $\Psi(t)=(W,X,Y,Z)$, which are obtained through a truncated Galerkin projection of the quasigeostrophic streamfunction $\psi(x,y,t)$ (for all details see Ref.\cite{pieriniJPO2011}). In the autonomous system subjected to a constant-in-time forcing with amplitude $\gamma$, the critical value $\gamma_c=1$ corresponds to a global bifurcation that separates an excitable system for $\gamma<\gamma_c$ (with small amplitude limit cycles) from a relaxation oscillator for $\gamma>\gamma_c$ (with limit cycles that include large amplitude ROs). 

\begin{figure}
\centering
\includegraphics[width=.35\textwidth,scale=.7]{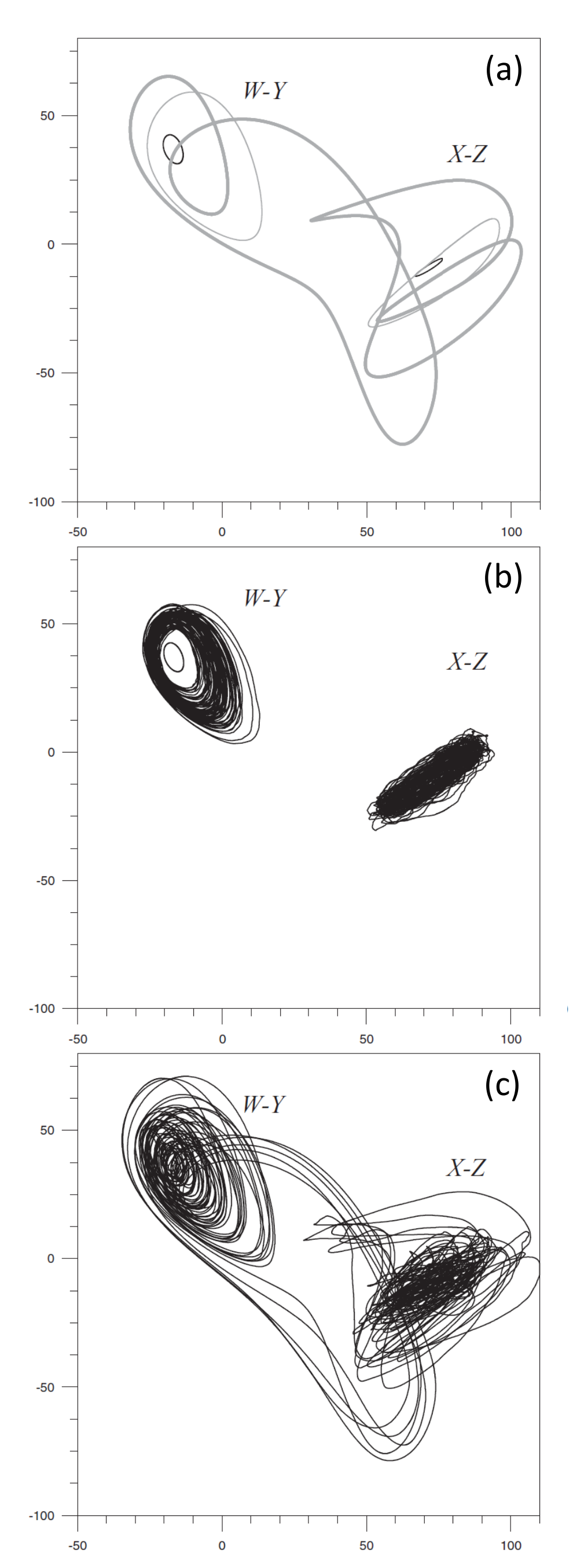}
\caption{(a): Limit cycles of the autonomous version of the excitable low-order model\cite{pieriniJPO2011} in the subcritical ($\gamma=0.94,0.97$, thin black and gray lines, respectively) and supercritical case ($\gamma=1.02$, thick gray line). (b): Evolution of the excitable system with $\gamma=0.94$ subjected to an additive red noise forcing with amplitude $\gamma_n=0.1$ and decorrelation time $T_d=0.1~yr$. (c): Same as panel (b) but with $T_d=2~yr$ (adapted from \citet{pieriniPR2012}).}
\label{figura_DE}
\end{figure}

Fig.~\ref{figura_DE}(a) (adapted from Ref.\cite{pieriniPR2012}) shows three stable limit cycles in the $(W,Y)$ and $(X,Z)$ planes: two of them lie in the excitable range ($\gamma=0.94,0.97$, thin black and gray lines, respectively) while the third one shows a typical RO ($\gamma=1.02$, thick gray line). Figs.~\ref{figura_DE}(b,c) show the effect, on the excitable system, of a time-dependent forcing in the form of an Orstein-Uhlenbeck noise with dimensionless amplitude $\gamma_n$ and decorrelation time $T_d$ (for a physical interpretation the interested reader should refer to\cite{pieriniPR2012}).

Fig.~\ref{figura_DE}(b) shows that, in the case $\gamma=0.94$ (i.e., in the subcritical excitable range) the forcing (with $\gamma_n=0.1$ and $T_d=0.1~yr$) deforms the limit cycle but is not able to excite the ROs. On the contrary, in Fig.~\ref{figura_DE}(c) (same $\gamma$ and $\gamma_n$ but $T_d=2~yr$), several ROs (5 in the specific forward time integration performed) are excited, whose basic features are preserved under variable forcing (compare the lines of Fig.~\ref{figura_DE}(c) with the thick gray lines of Fig.~\ref{figura_DE}(a)). In Ref.\cite{pieriniPR2012} a thorough analysis of the thresholds allowing excitation is presented not only for varying $T_d$ with constant $\gamma_n$ but also for varying $\gamma_n$ with constant $T_d$. 

The relevant properties worth stressing here are that (i) the excitation can occur only if a given control parameter crosses a certain threshold and (ii) the excitation modality has no effect during the RO evolution%(apart from some high-frequency variability)
, as only after a complete RO has elapsed can a new one be excited; in other words, a RO can be excited only if a reset time $\tau_{ro}$ equal to the temporal duration of the RO has elapsed after the time of excitation $t=t_p$ of the previous RO.

This example suggests that in a generic excitable system a RO is excited when the following conditions 1 and 2 are simultaneously satisfied:
\begin{enumerate}
\item a certain control variable $\zeta$ crosses a given threshold $\overline{\zeta}$ at time $\hat t$; 
\item $\hat t>\hat t_p+\tau_{ro}$, where $\hat t_p$ is the time at which the previous RO was triggered and $\tau_{ro}$ is the temporal duration of the RO.
\item $\dot S(\hat t)>0$
\end{enumerate}

\noindent  Condition 3 applies to the specific problem dealt with in Sect.~\ref{sec:res} and, although it seems obviously verified, it is actually required to rule out cases --that can occur in principle-- in which $\dot S(\hat t)<0$ with $\hat t=\hat t_p+\tau_{ro}$, i.e., as soon as the reset time has elapsed (condition 3 is in fact useless in the reference simulation of Sect.~\ref{sec:rs} but it is required for the sensitivity experiments of Sect.~\ref{sec:se} in which, in some cases, it does apply). 

In conclusion, the threshold crossing rule described by the three conditions above will be applied to the simulations presented in the next section to identify the last four glacial terminations of the LP. Intrinsic ROs of the climate system describing glacial cycles on a temporal scale of $\sim100$ kyr (Fig.~\ref{fig1}) will therefore be assumed.

Is there any modeling evidence that ROs of this kind exist in the climate system? The conceptual models of \citet{gildor-tzip00,gildor-tzip01} and \citet{pelletier03} suggest that this may be the case. For example, using an ocean-atmosphere-sea ice-land ice climate box model, \citet{gildor-tzip00,gildor-tzip01} showed that sea ice feedbacks can act as climate switches to produce ROs that account for the glacial-interglacial transitions on a $\sim100$ kyr time scale (with $\sim 30\%$ deviations for reasonable changes of parameter values) in the absence of any orbital forcing. Moreover, it was also shown that such ROs remain basically the same in the presence of seasonal and orbital variations in the solar radiation. Thus, these model results fit well within the RO assumption adopted in the present study.%, and connect this theoretical argument to a climate model that shares both qualitative and quantitative features of the glacial-interglacial variability as revealed by proxy records.

An important conceptual difference between the present approach and the studies\cite{gildor-tzip00,gildor-tzip01} is that, while in the latter the system is a relaxation oscillator (i.e., ROs are self-sustained), here the glacial climate is assumed to be in an excitable state, so that ROs need an external forcing to emerge; this allows the orbital forcing to fully manifest its role of pacemaker of the glacial cycles (Sect.~\ref{sec:res}). However, it can be conjectured that in those studies a suitable change in some control parameter would transform the relaxation oscillator into an excitable system without any substantial change in the character of the ROs: this would make that case consistent with the present approach. 

Finally, it is worth stressing that, while in the present approach the interglacials are assumed to terminate spontaneously, there is evidence that the orbital forcing may trigger abrupt changes in an interglacial \cite[e.g.,][]{ji_ea06,yin_ea21} associated with the rapid weakening of the Atlantic meridional overturning circulation; the end of an interglacial is, in any case,  followed by a slow return to a full glacial state driven by mechanisms internal to the climate system. Thus, even in this hybrid interpretation it can be plausibly assumed that a certain restoring time (close to $\tau_{ro}$) must be exceeded for a new interglacial to arise. The inclusion of this phenomenon in an extended version of the DE paradigm will be the subject of a future investigation.  
%@@@@@@@@@@@@@@@@@@@@@@@@@@@@@@@@@@@@@@@@@@@@@@@@
% ROSSO
%{\re } %@@@@@@@@@@@@@@@@@@@@@@@@@@@@@@@@@@@@@@@@@@@@@@@@ 

%====================================================
%====================================================
%====================================================
\section{\label{sec:res}Results}
%====================================================
%====================================================
\subsection{\label{sec:rs}The modeling strategy and the reference simulation}
The modeling strategy coherent with the aim of the present study (Sect.~\ref{sec:intro}) is the following. The fluctuations about a reference glacial state induced by the orbital forcing must first be obtained with a simple model, possibly described by a single ODE (thus, with just one prognostic variable, say $\zeta$) that lacks the property of excitability. Secondly, the threshold crossing rules defined in Sect.~\ref{sec:dep} must be applied to $\zeta$: they will parameterize the RO excitations and will therefore provide the timing of the simulated glacial terminations. The model and those rules, in combination, will provide an extremely idealized, "minimal" excitable glacial-cycle model. A good agreement between the timing thus obtained and that derived by proxy records would make this exercise useful despite --in fact, thanks to-- the simplicity of the approach.

Which prognostic variable would be more convenient to use? In the present context all variables describing global properties of the climate would be basically equivalent, because they would all yield virtually the same variability over the large temporal scales of interest, apart from small phase shifts and other differences that would not be significant here. Fig.~\ref{fig1} shows an example of this behavior referring to a record of benthic $\delta^{18}$O ocean sediments (which is a proxy of global ice volume and deep ocean temperature change) and to a global surface temperature record. The same would occur for the sea level, CO$_2$ content, etc. (e.g., as shown in several articles quoted in Sect.~\ref{sec:intro}).

Said this, the global surface temperature $T_s$ seems a good prognostic variable for two main reasons: (i) using $T_s$ is perfectly adhering to the Milankowitch hypothesis, which points to an increase of global temperature as a consequence of an increase in radiative forcing; (ii) for $T_s$ as the single prognostic variable one can rely on many simple energy balance models that have been --and still are-- among the simplest and yet most significant mathematical tools of theoretical climatology. 
\iffalse
\vspace{.7cm}
{\re qui spiegare perche' si usa Ts come prognostic variable e quindi perche' si usa un EBM

d18 (L-R) Ts, deep ocean temperature, sea level --and consequently ice-volume--, CO2 (Hansen), temperatures derived from ice cores (wunsch), ice,  (all with with some phase shift)
}
\fi

Thus, in the simulations the prognostic variable will be the Earth's surface temperature $\zeta \equiv T_s$ and the following energy balance model will be used: 
\begin{equation}\label{ebmo} 
		C_s\frac{dT_s}{dt}=\frac{S(t)}{4}\left [ 1-\alpha \left ( T_s \right ) \right ]-\tilde{\epsilon} \sigma T_s^{4},
\end{equation}
\noindent
where $S$ is the solar irradiance, $\alpha$ is the Earth's albedo, $\tilde{\epsilon}=1-\epsilon/2$ is a bulk emissivity that takes into account the greenhouse effect (where $\epsilon$ is the average emissivity of
the atmosphere in the infrared), $\sigma$ is the Stefan-Boltzmann constant and $C_s$ is the heat capacity of the Earth. The derivation of Eq.~\ref{ebmo} and a bifurcation analysis of the corresponding autonomous system are reported in Appendix A.

The definition of $S(t)$ is now discussed. The astronomical data used in this study are provided by the La2010 orbital solution of \citet[][]{laskar_ea11}, which represents an improvement in the adjustment of the parameters and initial conditions with respect to the previous La2004 solution \citep[][]{laskar_ea04}. 

\begin{figure}[ht]
\centering
\includegraphics[width=.45\textwidth,scale=.7]{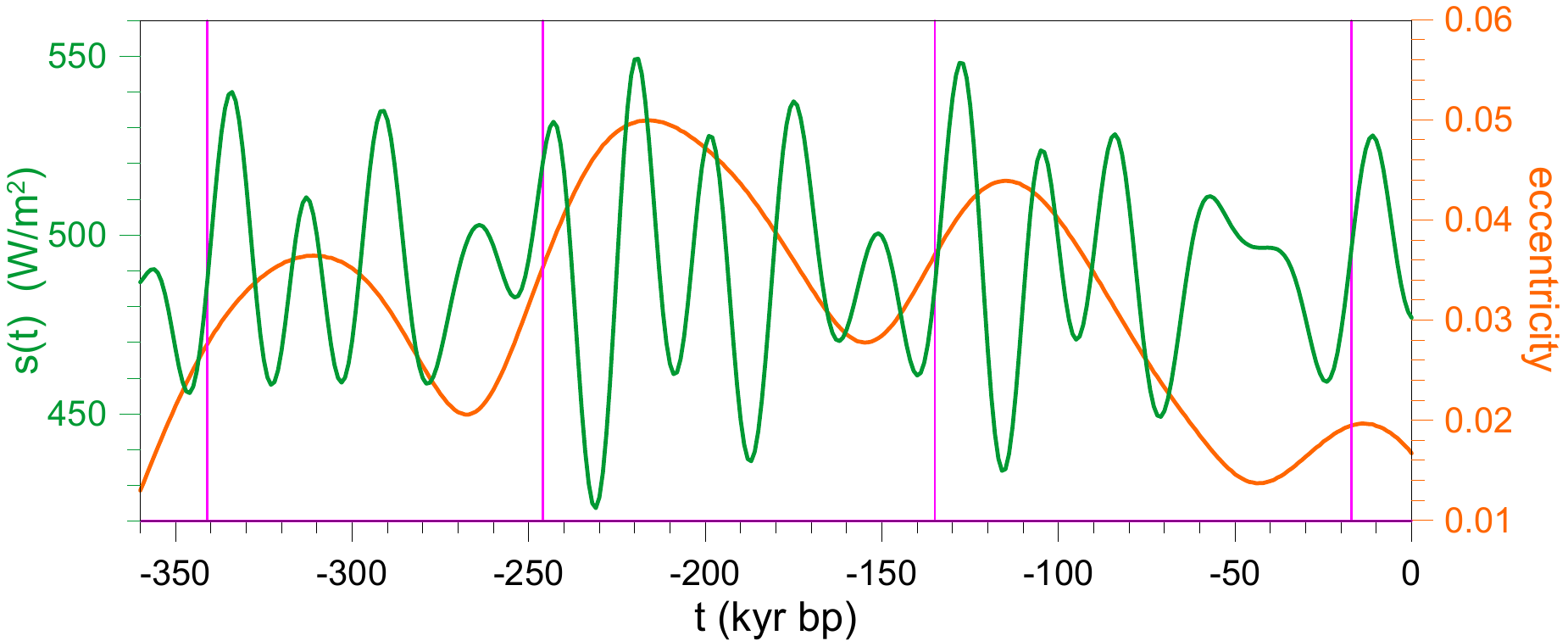}
\caption{Mean daily insolation at $65^{\circ}$N on summer solstice (green line) and Earth's eccentricity (orange line; both from \citet{laskar_ea11}). For the definition of the magenta vertical lines see Fig.~\ref{fig1}.}
\label{fig2}
\end{figure}

In an energy balance model, such as the one adopted here, the global annual mean insolation, basically dependent on the Earth's eccentricity $e$ (orange line of Fig.~\ref{fig2}) would seem to be the most obvious choice. However, in the present study, relying on the insolation at high northern latitudes is the correct alternative. In fact, it is mainly at those latitudes that changes in the incoming solar radiation control the melting and buildup of the northern ice sheets (e.g., the Laurentide and Fennoscandian ice sheets of the glacial periods) which, in turn, control the ice ages. 

For this reason we will use here, as often done in paleoclimate studies, the mean daily insolation at $65^{\circ}$N on summer solstice, whose time series $s\left( t \right)$ is shown by the green line in Fig.~\ref{fig2}. It is worth stressing that $s$ is affected by the Earth's axial tilt (obliquity) $\mathrm{\Phi}$, which controls the annual mean insolation at high latitudes and yields a dominant periodicity at about 41 kyr ($\mathrm{\Phi}$ is believed to be responsible for the timing of the glacial cycles before the middle Pleistocene transition, MPT). But $s$ depends also on the climatic precession $P=e\sin(\overline{\omega})$, where $\overline{\omega}$ is the longitude of perihelion from the moving equinox (yielding two main spectral peaks at 23 and 19 kyr).

To define the time-dependent solar irradiance $S$ appearing in Eq.~\ref{ebmo} by means of $s$, here the latter is first normalized,
\begin{equation}\label{insnorm}  
	s_n\left( t \right)=\frac{s\left( t \right)-\overline{s}}{s_{rms}}.
\end{equation}
The perturbation $\delta \cdot s_n\left( t \right)$ is then added to the solar constant, 
\begin{equation}\label{insola}  
	S\left( t \right)=S_{0}+\delta \cdot  s_n\left( t \right),
\end{equation}
where $\delta$ is a dimensional amplitude.   

Reasonably realistic parameter values will now be introduced. The reference temperature for an interglacial can be chosen as $T_{s(intergl)}=288^{\circ}K$: this value is found to correspond to a stable fixed point of the system with constant insolation ($\delta=0$ in Eq.~\ref{insola}) if $\epsilon=0.78$. A reference emissivity for a glacial state can instead be chosen as $\epsilon=0.72$, corresponding to a stable fixed point at $T_{s(glacial)}=284.1^{\circ}K$ (see the Appendix A), i.e., to a global temperature of about $4^{\circ}K$ less, in line with climate proxy estimates \cite[e.g.,][]{hansen_ea10,hansen_ea13,annan13,tierney_ea20}. In particular, the global surface temperature estimate from \citet{hansen_ea13} shown in Fig.~\ref{fig1} confirms the validity of this choice. Thus, $\epsilon=0.72$ is the value adopted in the simulations, which are aimed at identifying the times of termination of the glaciations. 

The albedo $\alpha$ is chosen to depend on $T_s$ as shown in Fig. A1 reported
in the Appendix A (this yields the existence of a stable snowball Earth state, as shown therein, but this is irrelevant in the present study).

As for $S$, the value of $\delta$ in Eq.~\ref{insola} deserves particular attention. As already pointed out, despite the energy-balance nature of the model, the eccentricity-dependent global annual mean insolation cannot be used as a radiative forcing while one should rather rely on the insolation shown by the green line in Fig.~\ref{fig2}; this includes the predominant effect of the axial tilt and climatic precession and, along with them, a wealth of additional periodicities. At the same time, one cannot use such a signal to force a model whose prognostic variable is the Earth's mean temperature. A compromise can be obtained by the hybrid approach proposed in Eq.~\ref{insola}. The value of $\delta$ therein is chosen in such a way that $S_{rms}$ is a given percentage of $S_0$. Here $\delta=6.41~Wm^{-2}$ corresponding to $S_{rms}=6.84~Wm^{-2}$, which is $0.5\%$ of the solar constant.

The remaining parameters to be defined are the two thresholds $\overline{T}_s=T_{s(glacial)}+\mathrm{\Delta}T$ and $\tau_{ro}$ according to the rule defined in Sect.~\ref{sec:dep}. The value $\mathrm{\Delta} T=0.5^{\circ}K$ is a reasonable choice; $\tau_{ro}=80$ kyr is also a reasonable value based on the character of the LP ice age (Fig.~\ref{fig1}). Note that these values of $\delta$, $\mathrm{\Delta} T$ and $\tau_{ro}$ will be varied in Sect.~\ref{sec:se}. 

\begin{figure}[ht]
	\centering
	\includegraphics[width=.45\textwidth,scale=.7]{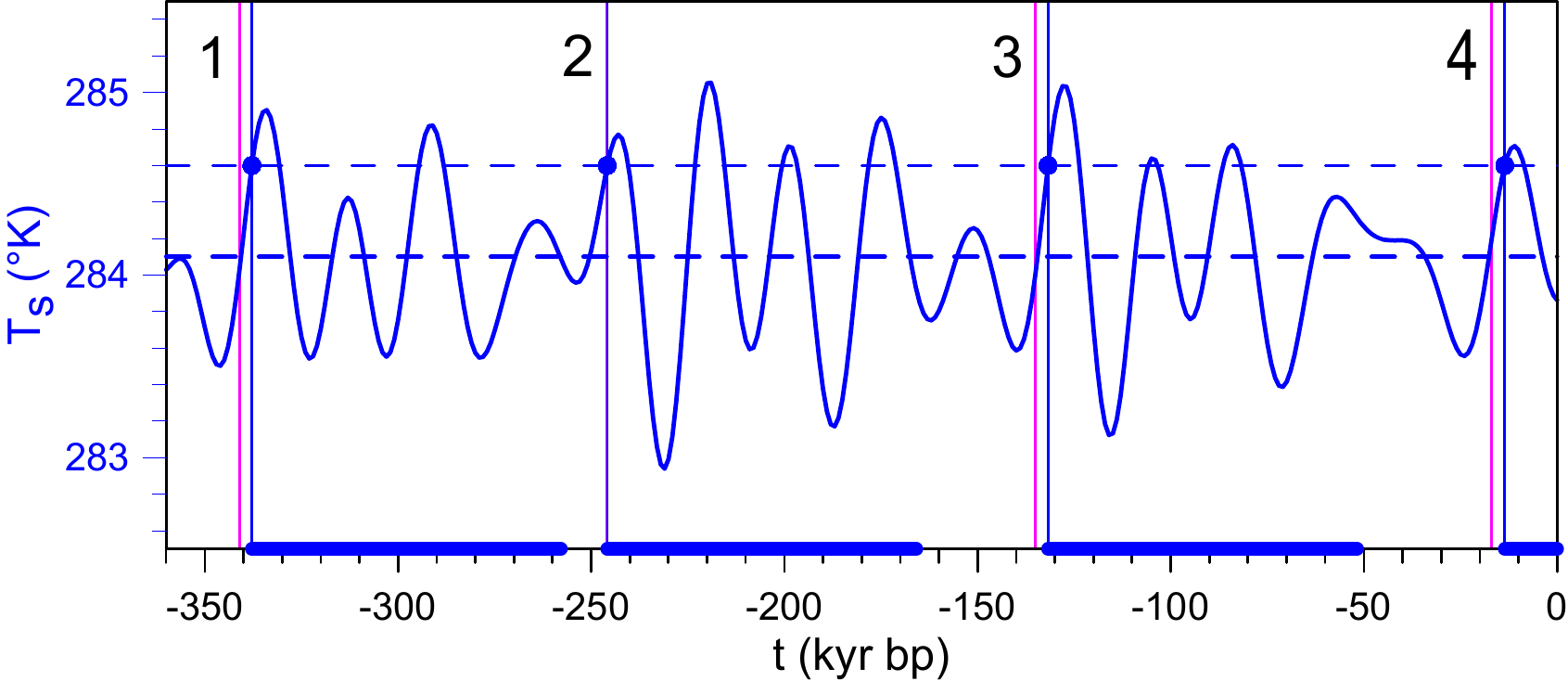}
	\caption{Timing of the simulated glacial terminations vs. timing derived from proxy data. Continuous blue line: $T_s(t)$. Thick dashed blue line: $T_{s(glacial)}$. Thin dashed blue line: $\overline{T}_s$. The thick blue segments have a temporal length $\tau_{ro}$ and begin at $\hat t_{i(mod)}$, i.e., when a glacial termination is identified by the model; the same times are also indicated by a dot and a vertical blue line. The vertical magenta lines indicate the times $\hat t_{i(proxy)}$ at which the glacial terminations have occurred according to proxy data (see Fig.~\ref{fig1}).
	}
	\label{fig3}
\end{figure}

Fig.~\ref{fig3} shows $T_s(t)$ in the interval of interest (from $-360$ kry bp to the present time) obtained by solving Eq.~\ref{ebmo}. The glacial termination times $\hat t_{i(mod)}$ ($i=1,...,4$) determined by the threshold crossing rules of Sect.~\ref{sec:dep} are indicated by the dots and vertical blue lines; their comparison with the times $\hat t_{i(proxy)}$ (again, $i=1,...,4$) of the real glacial terminations as determined by proxy data (magenta vertical lines, see Fig.~\ref{fig1}) shows excellent agreement.

It is worth noting that the $\tau_{ro}$-threshold plays a fundamental role in the identification of the transitions. For example, the $\overline{T}$-threshold is exceeded one or more times between to subsequent interglacials (once between the interglacials 1-2, three times between 2-3 and twice between 3-4), but this does not give rise to further excitations as it occurs before the complete relaxation of the hypothetical RO has come to an end; only afterwards can a new RO be excited. %On the other hand, during those intermediate time intervals the real temperature would be higher than the one modeled here, so allowing for premature excitations would be meaningless.

In conclusion, this simulation suggests that the simple DE paradigm defined by the threshold crossing rule of Sect.~\ref{sec:dep} provides a plausible theoretical basis for the explanation of the timing of the last four glacial termination of the LP within a purely deterministic framework. 

%====================================================
%====================================================
\subsection{\label{sec:se}Sensitivity experiments}
Numerical experiments are now presented with the aim of analyzing the sensitivity of the reference simulation with respect to the parameters $\mathrm{\Delta}T$, $\tau_{ro}$ and $\delta$. In each experiment one of these parameters is varied while all the others are unchanged. For each value of the variable parameter, the glacial termination times $\hat t_{i(mod)}$ ($i=1,...,N_m$) are computed and the $N_f\le 4$ of them that fit one of the four proxy times $\hat t_{k(proxy)}$ are identified. $\hat t_{i(mod)}$ is assumed to fit $\hat t_{k(proxy)}$ if $\left| \hat t_{i(mod)}-\hat t_{k(proxy)}  \right|\le d_{0}$, where $d_0=10$ kyr, considered as a sufficiently small temporal departure.

\begin{figure}[ht]
	\centering
	\includegraphics[width=.45\textwidth,scale=.7]{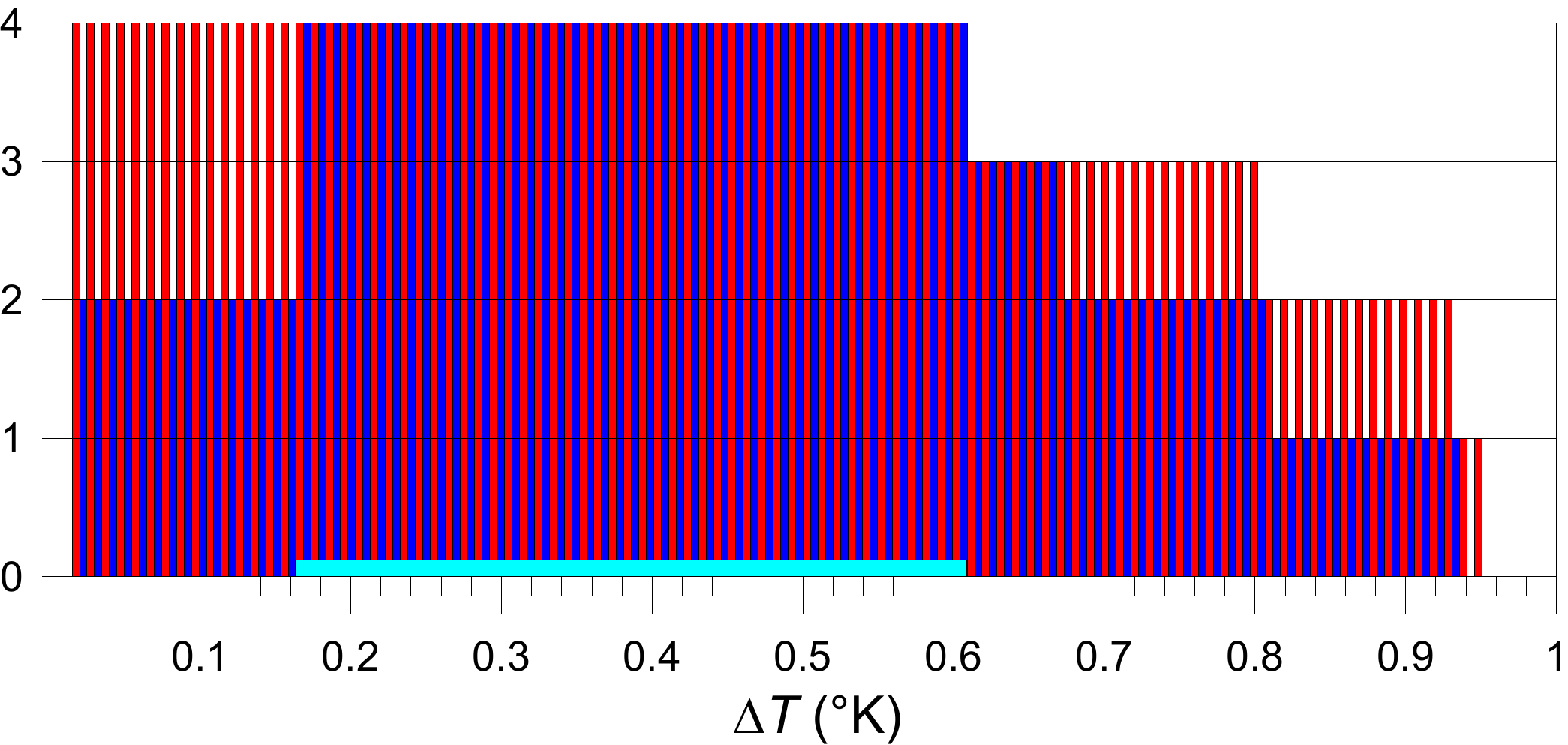}
	\caption{Red bars: number $N_m$ of simulated glacial termination times $\hat t_{i(mod)}$ as a function of $\mathrm{\Delta}T$ (all the other parameters of the reference simulation are unchanged). Blue bars: number $N_f$ of simulated termination times that fit one of the proxy times $\hat t_{k(proxy)}$. The cyan bar shows the $\mathrm{\Delta}T$-range in which all (four) the simulated transition times fit all (four) the real transition times, thus producing a timing that is basically the same as that shown in Fig.~\ref{fig3}. 
	}
	\label{fig4}
\end{figure}

Fig.~\ref{fig4} shows $N_m$ (red bars) and $N_f$ (blue bars) as functions of $\mathrm{\Delta}T$. For $\mathrm{\Delta}T\lesssim 0.16^{\circ}K$, $N_m=4$ but $N_f=2$. This is because $\hat t_{3(mod)}$ does not fit $\hat t_{3(proxy)}$, being selected just before the small $T_s$--peak at $t\cong -155$ kyr (see Fig.~\ref{fig3}); analogously, $\hat t_{4(mod)}$ does not fit $\hat t_{4(proxy)}$ (the Holocene inception) being selected just before the small $T_s$--peak at $t\cong -60$ kyr. For $0.6 \lesssim \mathrm{\Delta}T \lesssim 0.675^{\circ}K$, the Holocene inception is missed because the $T_s$--peak at $t\cong -10$ kyr lies below the $\overline{T}$-threshold. For higher $\mathrm{\Delta}T$ other terminations are missed. Transition times that are basically the same as those shown in Fig.~\ref{fig3} are obtained in the range $T_s\cong \left( 0.16, 0.6 \right)^{\circ}K$ (evidenced by the cyan bar in the graph).

\begin{figure}[ht]
	\centering
	\includegraphics[width=.45\textwidth,scale=.7]{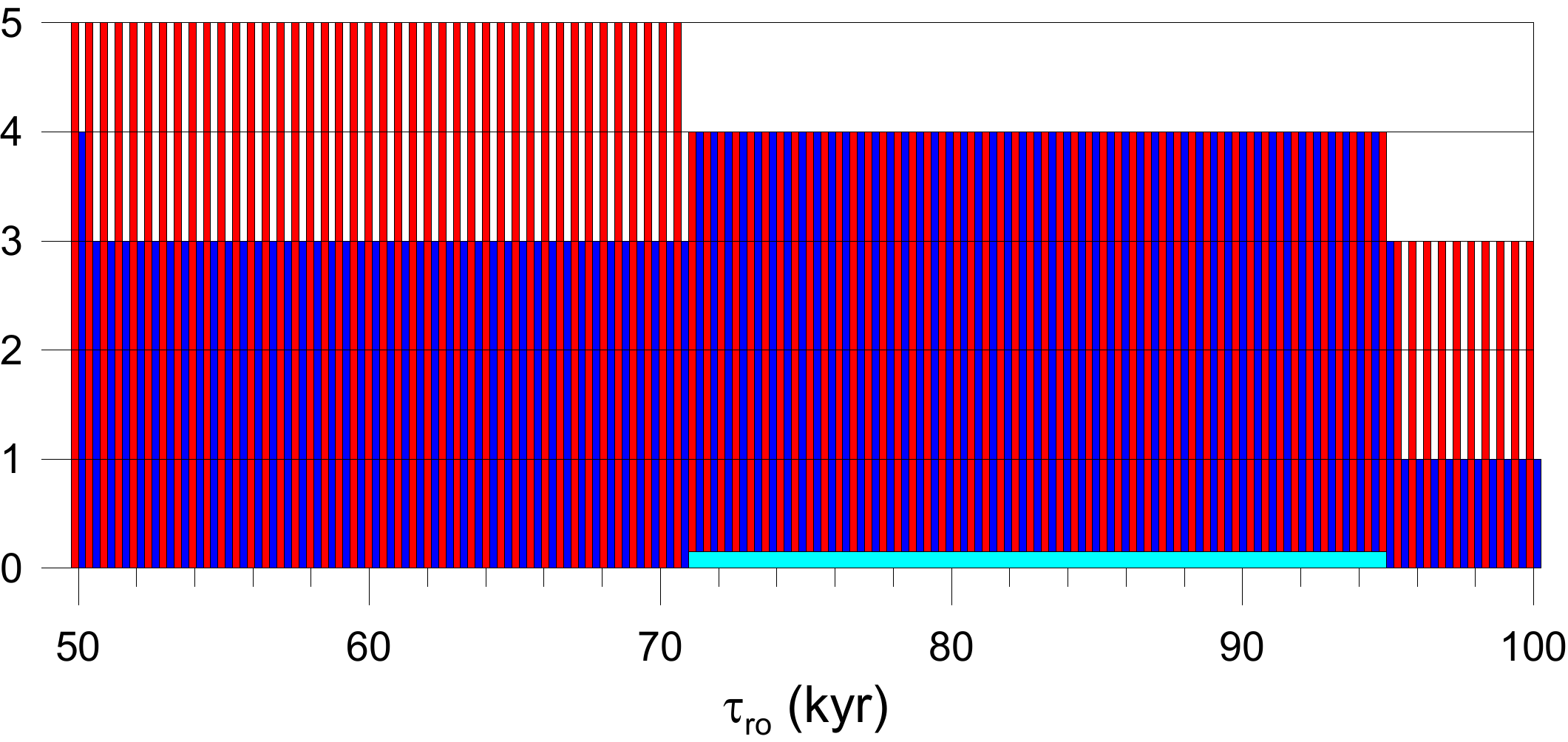}
	\caption{Same as Fig.~\ref{fig4} but for $N_m$ and $N_f$ as functions of $\tau_{ro}$.
	}
	\label{fig5}
\end{figure}

Fig.~\ref{fig5} shows $N_m$ and $N_f$ as functions of $\tau_{ro}$. For $10 \le \tau_{ro} \le 50$ kyr (not shown) all the four real terminations are captured but other excitations (up to 6) occur as well.  For $50 < \tau_{ro} \lesssim 70$ kyr, $N_m=5$ but $N_f=3$: this is because transition 3 is preceded by a glacial termination just before the $T_s$--peak at $t\cong -175$ kyr (Fig.~\ref{fig3}). Transition times that are basically the same as those shown in Fig.~\ref{fig3} are obtained in the range $\tau_{ro}\cong \left( 71, 95 \right)$ kyr. For larger $\tau_{ro}$, after the identification of transition 1 the other three real transitions cannot be captured, being not allowed by the $\tau_{ro}$-threshold, but two transitions far from any real one are identified (so $N_m=3$ and $N_f=1$). 

\begin{figure}[ht]
	\centering
	\includegraphics[width=.45\textwidth,scale=.7]{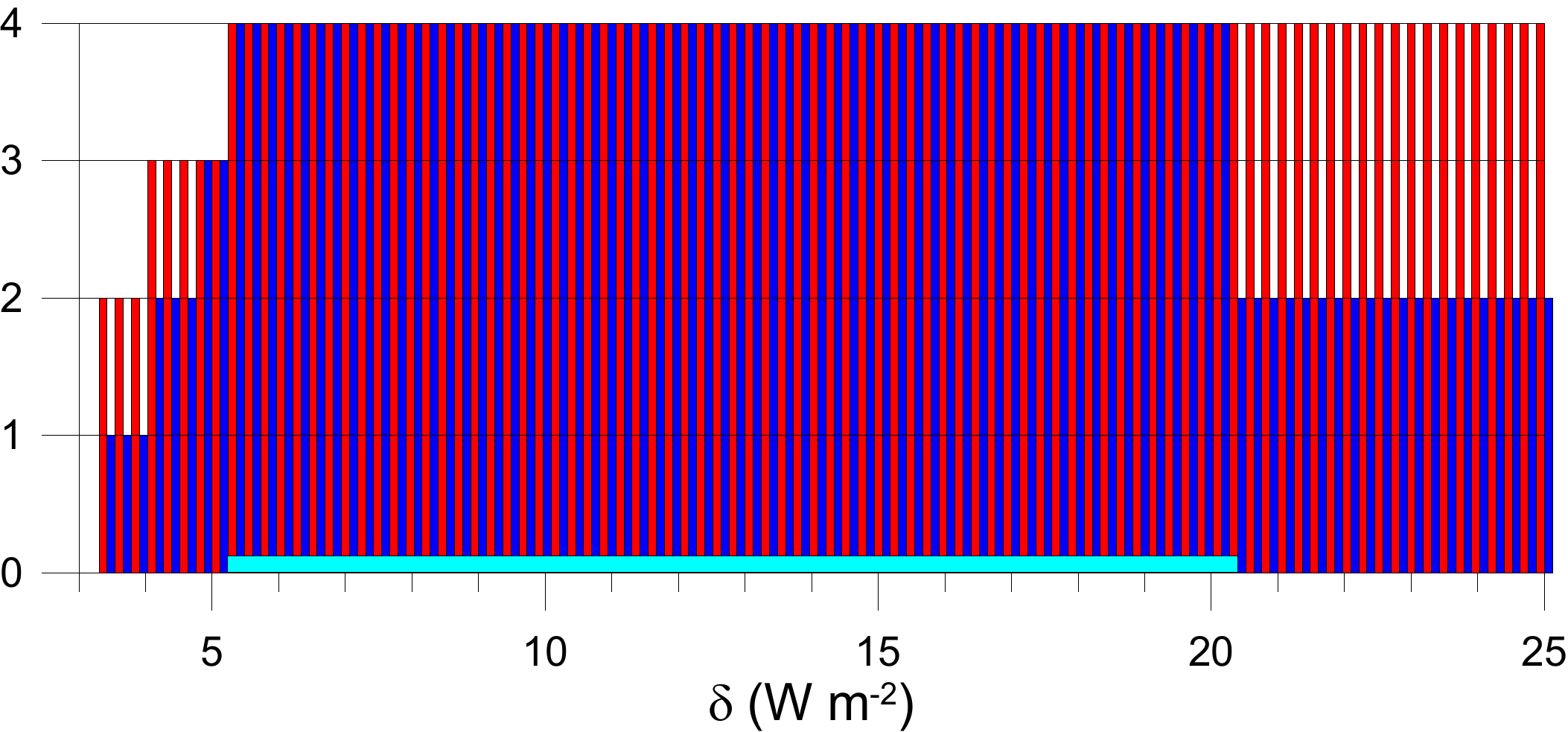}
	\caption{Same as Fig.~\ref{fig4} but for $N_m$ and $N_f$ as functions of $\delta$.
	}
	\label{fig6}
\end{figure}

Finally, Fig.~\ref{fig6} shows $N_m$ and $N_f$ as functions of $\delta$. For $\delta \gtrsim 5~Wm^{-2}$ the forcing amplitude is sufficient to reproduce correctly all the four glacial terminations; in fact, transition times that are basically the same as those shown in Fig.~\ref{fig3} are obtained in the range $\delta\cong \left( 5, 20 \right)Wm^{-2}$. For $\delta \gtrsim 20~Wm^{-2}$  transitions 1 and 2 are correctly captured, while transitions 3 and 4 are preceded by two spurious terminations just before the $T_s$--peaks at $t\cong -155$ kyr and  $t\cong -60$ kyr, respectively (so $N_m=4$ and $N_f=2$).

In conclusion, the correct timing of the four glacial terminations of the LP reproduced by the reference simulation and shown in Fig.~\ref{fig3} is very robust, as it is preserved if $\mathrm{\Delta}T$, $\tau_{ro}$ and $\delta$ are varied within the wide ranges shown by the cyan bars in Figs.~\ref{fig4},\ref{fig5},\ref{fig6}. 

%{\re ccc }
 
%====================================================
%====================================================
%====================================================
\section{\label{sec:disc}Further discussion}

%====================================================
%====================================================
\subsection{\label{sec:noise}On the role of noise}
In the previous section, it was shown that the DE paradigm %to an energy balance model, describing the temperature fluctuations about a reference glacial state, 
can provide a correct and robust timing of the LP glacial terminations in a purely deterministic framework, in line with other conceptual model results (see Sect.~\ref{sec:intro}). On the contrary, in models based on the stochastic resonance (SR) mechanism (e.g., \citet{benzi_ea82,nicolis82,gammaitoni98}, etc.) noise is claimed to play a fundamental role in the glacial-interglacial variability. Thus, it is useful to briefly analyze the reasons for this apparent contradiction.

SR is a fascinating dynamical mechanism which was recently invoked, among other important scientific discoveries, in connection with the award of the 2021 Nobel Prize in Physics\cite{dedomenico-vulpiani21}. In general, a suitable noise is known to induce transitions between two stable climate states \cite{sutera81,nicolisx281}. In SR a strong white noise and a weak periodic forcing cooperate in such a way that noise-induced transitions between two stable states (the glacial-interglacial transition and vice versa) occur with the same period of the forcing: for this to happen, the noise amplitude and the average noise-induced switching time must satisfy a certain matching condition. The basic idea is that the periodicity of the forcing should be reflected into the 100-kyr quasiperiodicity of the LP climate variability, so in SR the eccentricity $e(t)$ was invoked. 

The main criticism leveled against this approach was that in SR glacial inceptions occur abruptly, as well as the glacial terminations, in sharp contrast with the typical saw-tooth shape evidenced in the paleorecords (Fig.~\ref{fig1}). However, there are also problems concerning the orbital forcing that are worth considering. As shown in Fig.~\ref{fig2}, $e(t)$ is far from being a simple periodic signal (it yields spectral lines at 95, 100, 123, 131 kyr, e.g., \citet{riechers_ea22}, so that actually no significant spectral peak at 100 kyr is present in the orbital forcing); but the periodicity of the forcing is a prerequisite for SR to be applied. Furthermore, $e$ has a minor effect on the insolation at high northern latitudes (the one that matters, see Sect.~\ref{sec:rs}); so, $e$ cannot be responsible for pacing the glacial terminations. Thus, the deterministic forcing component of SR does not meet the conditions required by this paradigm to be applied with realistic orbital forcing.

But more in general, even if the previous inconsistencies were absent, since the matching condition required by SR cannot be accidentally verified (it could be so for one cycle but hardly for several consecutive cycles), a peculiar dynamical mechanism connecting very different time scales should be in play to grant such condition; but no mechanism of this kind has ever been proposed in the context of the ice ages. In conclusion, the SR mechanism is simply inapplicable to the glacial-interglacial variability problem (but it is, on the contrary, of great interest in many other fields of nonlinear science, e.g.,\cite{gammaitoni98,gammaitoni09}). This shows, more in general, how unlikely is the possibility of a stochastically controlled quasiperiodicity in climate dynamics.

On the other hand, stochasticity in climate models can also have the classical effect of introducing random deviations from a regular behavior\cite{imkvst01,palmer_williams10}. More specifically, climate models perturbed by noise can experience relevant changes in the timing of the glacial terminations and, more in general, even qualitative changes due to the presence of bifurcations and chaos\cite{alexandrov_ea21,riechers_ea22}, and this can occur also in threshold crossing models. For example, in a two-threshold model of the glacial cycles preceding the MPT, \citet{berends_ea21} reported a change from deterministic to chaotic timing of the glacial terminations if a small amount of white noise is added to an idealized sinusoidal insolation.

The question therefore arises as to why noise may not be effective in pacing the glacial cycles by the orbital forcing. In their analysis of the nonlinear phase locking of the glacial cycles to Milankovitch forcing, \citet{tzip_ea22} showed that the latter could be setting the phase of those cycles even in the presence of abundant noise in the climate system. Besides, the ice-sheet evolution is less noisy than the atmosphere and even the ocean dynamics, since it has a long memory of previous ice accumulation or melting. Therefore, in the glacial cycles the effects of internal climate variability parameterizable with noise are expected to be less relevant than in other climate phenomena. This is consistent with a pacing of the cycle by the mere orbital forcing, and so with the original Milankovitch hypothesis.

In a future investigation, the threshold crossing rules defined in Sect.~\ref{sec:dep} and applied in Sect.~\ref{sec:res} will be extended to include noise. This will allow to derive useful information about the character of the noise that preserves the right timing of the glacial terminations and, so, the validity of the DE paradigm even in the presence of internal climate fluctuations.

%{\re %@@@@@@@@@@@@@@@@@@@@@@@@@@@@@@@@@@@@@@@@@@@@@@@ }
%====================================================
%====================================================
\subsection{\label{sec:cp}On the 100-kyr cycle problem}
%{\dgr in the power spectra the phase is lost but this is fundamental to determine the total amplitude of the forcing, which is the only parameter that matters in the DE paradigm}
The present conceptual model suggests that, comparing the power spectral densities of paleorecords with those of the orbital forcing (a typical approach in climate dynamics) may be problematic for those climatic phenomena (if any) that are ruled by DE. This is because in the DE mechanism the triggering of a RO (corresponding to a glacial termination) depends only on the instantaneous value of a given control parameter ($T_s$ in the present case) as it fluctuates about the glacial state, representing a virtually linear response to the \textit{instantaneous total} orbital forcing. But the latter depends, not only on the amplitude of each Fourier component, but also on its phase, which is lost in the power spectrum.   

In this context, insofar as the simple DE paradigm applies to the ice ages, one of the basic quandaries of the Quaternary glaciation cycles \cite{imbrie_ea93,huybers07,riechers_ea22} could be resolved as follows. The question is: why is there a climate's spectral peak at $\sim100 $ kyr given that no significant analogous peak is present in the forcing? The answer may formulated as follows. Intrinsic ROs describing a glacial cycle with a relaxation time $\tau_{ro}\simeq70\div95$ kyr (Fig.~\ref{fig5}) are assumed to exist in the climate system \cite{gildor-tzip01}. Moreover, the small temporal scales ($\sim20 $ kyr) of the climatic precession --as modulated by both obliquity and eccentricity (green line in Fig.~\ref{fig2})-- may trigger a  RO only few tens of kyr after the end of a previous RO. This makes the temporal distance between two successive glacial terminations to be around, or just above $100$ kyr, as demonstrated in Fig.~\ref{fig3}.   

So, paradoxically, according to this interpretation, the highest frequency components of the orbital forcing (included in the climatic precession) are those that would give an essential contribution to the observed $\sim 100$ kyr cycles of the LP.  

%====================================================
%====================================================
%====================================================
\section{\label{sec:conc}Summary and conclusions}
In this paper a deterministic dynamical paradigm (deterministic excitation, DE) has been formulated, and successfully tested, with the aim of characterizing in one of the simplest possible ways the link between orbital forcing and glacial-interglacial transitions in the late Pleistocene (LP), as implied by the Milankovitch hypothesis.

First, the DE paradigm is introduced. To do this, the concept of relaxation oscillation (RO) is recalled and elucidated by means of a low-order model of the wind-driven ocean circulation. For an operational definition of DE, threshold crossing rules parameterizing the activation of internal climate feedbacks leading to a RO excitation are then defined. According to the DE paradigm, a glacial-interglacial transition requires the following conditions to occur:

\begin{itemize}
	\item[$\bullet$] The climate must be an excitable system lying in a stable glacial state and must possess nonlinear intrinsic ROs --describing glacial-interglacial cycles-- with a temporal scale $\tau_{ro}$ that, in the LP, is $\tau_{ro}\sim 100$ kyr. The ROs (which are not self-sustained, i.e., they do not emerge spontaneously) can be excited by the orbital forcing. The time of excitation $\hat t$ of a RO corresponds to the beginning of a glacial-interglacial transition.	
	\item[$\bullet$] $\hat t$ must exceed $\hat t_p+\tau_{ro}$, where $\hat t_p$ is the time at which the previous RO was triggered (in other words, the temporal distance between two successive excitations cannot be less than $\tau_{ro}$). 
	\item[$\bullet$] At $t=\hat t$, a control parameter $\zeta$ must exceed a given \textit{tipping point} (or threshold) $\overline{\zeta}$: $\zeta(t<\hat t)<\overline{\zeta}$; $\zeta(t\ge\hat t)\ge\overline{\zeta}$	
\end{itemize} 

Secondly, an energy balance model subjected to a forcing representing the solar radiation received in the summer at high northern latitudes, is used to describe the fluctuations induced by the orbital forcing on the glacial state, in the absence of any stochastic parameterization. This model (with $\zeta \equiv T_s$, where $T_s$ is the Earth's surface temperature) and the above mentioned rules provide, in combination, a minimal excitable deterministic glacial-cycle model. 

By applying the model and those rules with plausible parameter values in a reference simulation, the timing of the last four LP transitions is found to be in good agreement with proxy records. Moreover, sensitivity numerical experiments in which $\overline{T}_s$, $\tau_{ro}$ and the amplitude of the time-dependent forcing component $\delta$ are varied, show that the correct timing is preserved in wide ranges of parameter values.

These results show, in a particularly idealized and intuitive framework, that the correct and robust timing of the LP glacial terminations can be obtained in a purely deterministic context, in line with other conceptual model results. However, stochastic theories (such as the stochastic resonance) have been proposed in which noise is claimed to play a decisive role in the timing of the deglaciations. The reasons why a stochastically controlled quasiperiodicity in climate dynamics is unlikely is therefore discussed. 

Finally, it is shown how the DE paradigm might provide a possible explanation of the classical 100-kyr cycle quandary, which points to the contradiction between the existence of a spectral peak at around $100$ kyr in the proxy records and the absence of an analogous significant spectral peak in the orbital forcing. It is suggested that the typical temporal length $\tau_{ro}$ of the RO and the small time scale of the climatic precession provide, in combination, a temporal distance between two successive glacial terminations to be around, or just above 100 kyr, like in the proxy records.

In conclusion, thanks to its simplicity and to its ability to produce a realistic glacial termination timing in the LP, the DE paradigm may provide a plausible conceptual basis for the explanation of the glacial-interglacial variability and can serve as a reference guideline for climate models at different levels of complexity.

In a future study the DE paradigm will be extended to take into account variations of $\tau_{ro}$ due to changes in climate conditions and the effect of the orbital forcing on the interglacial terminations. Moreover, noise will be added to the model to obtain information about the character of the internal climate fluctuations that preserve the right timing of the glacial terminations and, so, the validity of the DE paradigm.

%---------------------------------------
%\vspace{.5cm}
\iffalse
{\dgr
PUNTO 55 DI TZIPPERMAN ET AL 2006: These findings suggest that "the ice ages problem is effectively divided into two separate sub problems: the first is explaining the phase or timing of the cycles, and the second is finding the physical mechanism that gives rise to these cycles." We believe that nonlinear phase locking provides a good framework for understanding the first problem, even if the second is still far from being resolved.

dire quali sono gli improvements che possono essere apportati alla DE

dire che un'eccitazione è come il superamento di un tipping point

In a future perspective, the present conceptual model will be extended to include ROs with a $\sim100$-kyr time scale and noise to test the hypothesis mentioned above. The system's pullback attractor \cite[e.g.,][]{GCS.2008,drotos2015,PGC16,tel19,ghil-luc20} will also be estimated to analyze the interplay between the internal climate variability and the effect of the orbital forcing. Cases of generalized synchronization \cite[e.g.,][]{de2013astronomical,pieriniJPO2014,pg21} will as well be analyzed. 
}
\fi
%---------------------------------------

\begin{acknowledgments}
The author is glad to acknowledge the use of the Milankovitch orbital data of \citet{laskar_ea04,laskar_ea11} available at https://biocycle.atmos.colostate.edu/shiny/Milankovitch/, of the LR04 benthic $\delta^{18}$O stack of \citet{lisiecki05} available at https://lorraine-lisiecki.com/LR04stack.txt, and of the global surface temperature estimate of \citet{hansen_ea13} available in the respective supplementary material. The author is also glad to acknowledge the use of REV\TeX{} and Overleaf (https://www.overleaf.com/contact).

%Magenta line: LR04 benthic $\delta^{18}$O stack constructed by the graphic correlation of 57 globally distributed benthic $\delta^{18}$O records (from \citet{lisiecki05}; the magenta vertical lines mark the onset of the abrupt glacial-interglatial transitions). Blue line: global surface temperature estimate (from \citet{hansen_ea13}).

%the support of the author community in using REV\TeX{}, offering suggestions and encouragement, testing new versions,
%\dots.
\end{acknowledgments}

\section*{AUTHOR DECLARATIONS}
\subsection*{Conflict of Interest}
The author has no conflicts to disclose.
    
\section*{Data Availability Statement}
The data that support the findings of this study are available
from the corresponding author upon reasonable request.

% Data sharing is not applicable to this article as no new data were created or analyzed in this study
%AIP Publishing believes that all datasets underlying the conclusions of the paper should be available to readers. Authors are encouraged to deposit their datasets in publicly available repositories or present them in the main manuscript. All research articles must include a data availability statement stating where the data can be found. In this section, authors should add the respective statement from the chart below based on the availability of data in their paper.

%====================================================
%====================================================
%====================================================
%\appendix

\section*{APPENDIX A: THE ENERGY BALANCE MODEL}
\label{sec:app}
\renewcommand{\figurename}{Fig. A}
\setcounter{figure}{0}
\setcounter{equation}{0}
In this appendix the energy balance model is presented. Let us recall that the solar constant $S_0=1367.6~Wm^{-2}$ gives the mean total solar irradiance per unit area at a distance of one astronomical unit; thus, $\pi R^2 S_0$ ($R$ is the Earth's radius) is the mean  energy reaching the Earth's upper atmosphere and $S_0/4$ is the corresponding incoming radiation density. The energy density of the net absorbed radiation is therefore $S_0(1-\alpha)/4$, where  $\alpha$ is the Earth's albedo. 

If the Earth were a perfect blackbody, the following global energy balance would be satisfied:
\begin{equation}\label{balance1}\tag{A1}  
	\frac{S_0}{4}\left ( 1-\alpha \right )=\sigma T_e^{4}
\end{equation}
where $\sigma=5.67\times10^{-8}~Wm^{-2}K^{-4}$ is the Stefan-Boltzmann constant and $T_e$ is the mean Earth's temperature. The RHS of Eq.~\ref{balance1} can be modified by introducing the mean temperature of the Earth's surface $T_s$ and that of the atmosphere $T_a$:
\begin{equation}\label{balance2}\tag{A2}  
\frac{S_0}{4}\left ( 1-\alpha \right )=\epsilon \sigma T_a^{4}+(1-\epsilon)\sigma T_s^{4}
\end{equation}
\noindent 
where the RHS represents the net flux density out of the top of the atmosphere. Here the Earth's surface is assumed to emit as a perfect blackbody in the infrared (the term $\sigma T_s^4$) while the atmosphere is assumed to be transparent to the shortwave solar radiation and to emit a fraction $\epsilon$ of the blackbody radiation with temperature $T_a$ (the term $\epsilon \sigma T_a^4$) due to the presence of greenhouse gases, where $\epsilon$ is the average emissivity of the atmosphere in the infrared. Finally, the term $-\epsilon \sigma T_s^4$ represents the fraction of the upward radiation from the surface that is absorbed by the atmosphere (note that the absorptivity is equal to the emissivity according to the Kirchoff's law of thermal radiation).

The atmospheric energy balance requires that the energy flux emitted from the surface that is absorbed by the atmosphere be equal to that emitted by the atmosphere: 
\begin{equation}\label{balance3}\tag{A3}  
\epsilon \sigma T_s^{4}=2\epsilon \sigma T_a^{4}
\end{equation}
(due to the rapid atmospheric adjustment, such balance can be considered as instantaneously verified in our long-term application, bringing $T_a=2^{-1/4}T_s$). Note that the factor 2 in the RHS accounts for the radiation, in this idealized slab atmosphere, both up to space and down to the surface. 

By combining Eq.~\ref{balance2} with \ref{balance3} one gets:
\begin{equation}\label{balance5}\tag{A4} 
	\frac{S_0}{4}\left ( 1-\alpha \right )=\tilde{\epsilon} \sigma T_s^{4}
\end{equation}
\noindent
with $\tilde{\epsilon}=1-\epsilon/2$. Eq.~\ref{balance5} improves Eq.~\ref{balance1} in that it includes a better defined temperature and a bulk emissivity that explicitly takes into account the greenhouse effect.

\begin{figure}[ht]
	\centering
	\includegraphics[width=.4\textwidth,scale=.7]{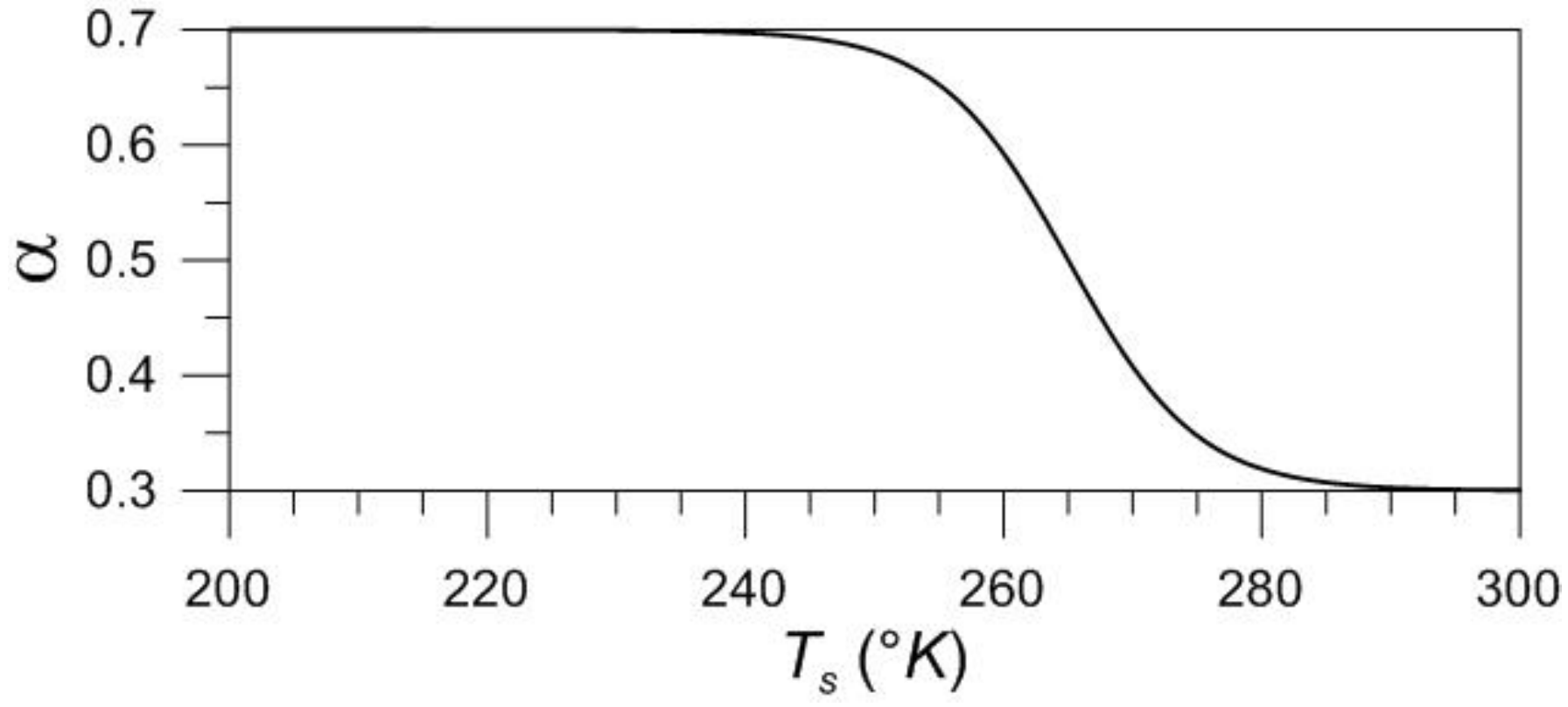}
	\caption{Temperature-dependent albedo (from \citet{flath12}).
	}
	\label{alfa}
\end{figure}

If the perfect energy balance in Eqs.~\ref{balance5} is relaxed, the evolution equation for $T_s$ is finally obtained: 
\begin{equation}\label{lom}\tag{A5} 
		C_s\frac{dT_s}{dt}=\frac{S(t)}{4}\left [ 1-\alpha \left ( T_s \right ) \right ]-\tilde{\epsilon} \sigma T_s^{4}.
\end{equation}
\noindent
where the heat capacity $C_s$ is chosen as $C_s=10^8J~m^{-2}K^{-1}$. The time-dependent solar irradiance $S(t)$ is defined in Sect.~\ref{sec:rs}. In addition, the temperature-dependent albedo is defined according to \citet[][]{flath12} and is shown in Fig.~A\ref{alfa}: $\alpha$ changes gradually from a value $\alpha \simeq 0.3$ typical of the LP to a value $\alpha \simeq 0.7$ appropriate for an ice-covered Earth for surface temperatures $\sim 40^{\circ}C$ smaller.

\begin{figure}[ht]
	\centering
	\includegraphics[width=.35\textwidth,scale=.7]{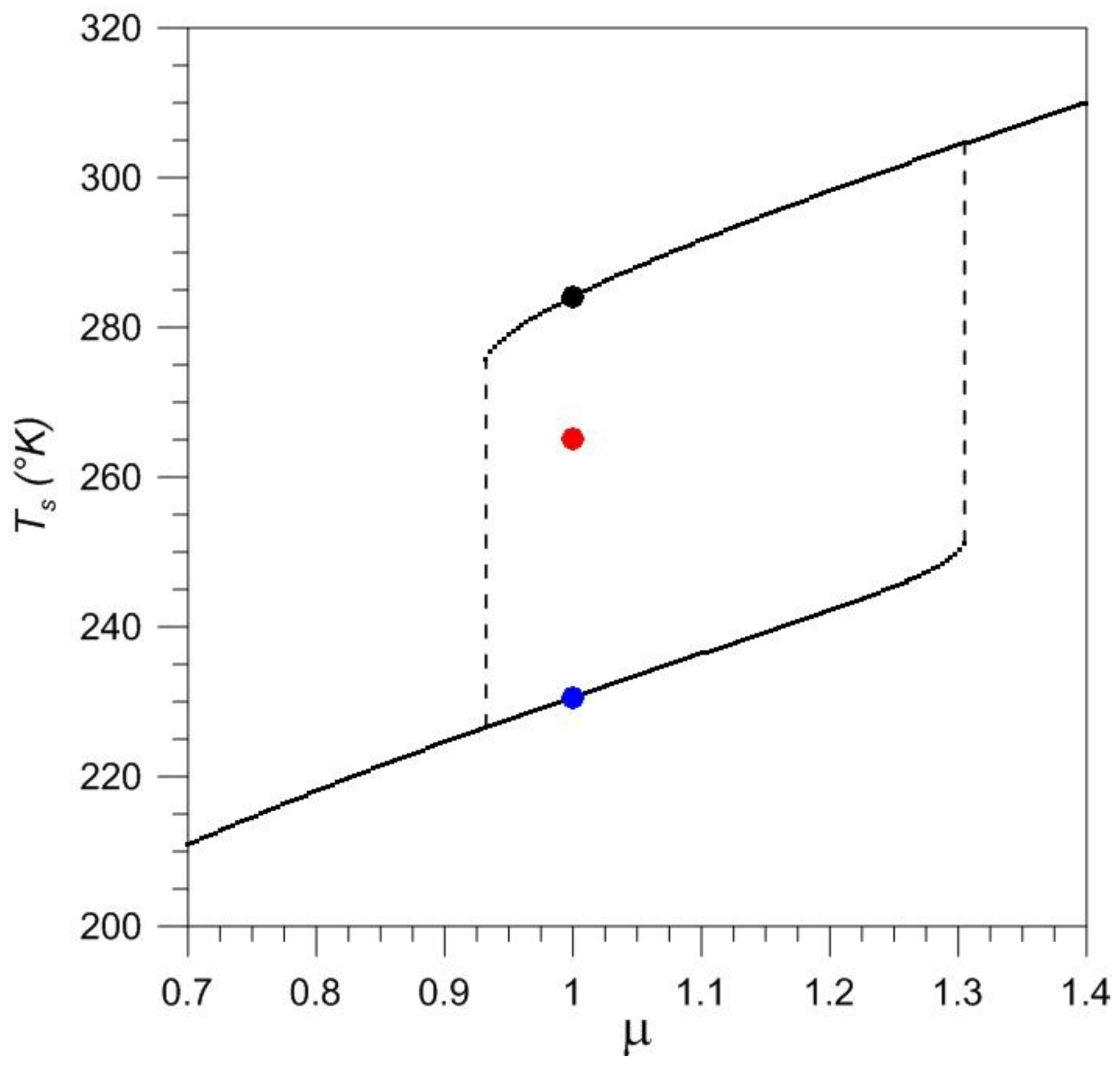}
	\caption{Bifurcation diagram with $T_s$ as a function of $\mu=S/S_0$ in the autonomous case. Small black dots: asymptotic state of $40,000$ points in the $(T_s,\mu)$ plane initially and uniformly distributed in the intervals $T_s: \left [ 200^{\circ}K,300^{\circ}K\right ]$; $\mu: \left [ 0.7,1.4 \right ]$. Big black dot: stable fixed point ($\mu=1,~T_s=284.1^{\circ}K$) corresponding to the glacial state of the reference simulation. Big blue dot: stable fixed point ($\mu=1,~T_s=230.6^{\circ}K$) corresponding to a snowball Earth state. Big red dot: unstable fixed point ($\mu=1,~T_s=265^{\circ}K$).
	}
	\label{bifurc}
\end{figure}

A bifurcation analysis of the autonomous system is now presented. The bifurcation diagram with $T_s$ as a function of the control parameter $\mu=S/S_0$, with constant $S$ in Eqs.~\ref{lom} (autonomous system) and with the settings of Sect.~\ref{sec:rs}, is reported in Fig.~A\ref{bifurc}. The diagram is obtained by computing numerically the asymptotic state of $40,000$ ($200\times 200$) points in the $(T_s,\mu)$ plane initially and uniformly distributed in the intervals $T_s: \left [ 200^{\circ}K,300^{\circ}K\right ]$; $\mu: \left [ 0.7,1.4 \right ]$. 

For $\mu=1$  a stable fixed point is present at $T_s=284.1^{\circ}K$ (big black dot in Fig.~A\ref{bifurc}, thick dashed blue line in Fig.~\ref{fig3}) corresponding to the LP glacial state. Still for $\mu=1$ two more fixed points are present: an unstable one at $T_s=265^{\circ}K$ (red dot) and a stable one at $T_s=230.6^{\circ}K$ (blue dot). This equilibrium point represents a snowball Earth state, extensively studied in the framework of energy balance models \cite[e.g.,][]{budyko69,sellers69,ghil76,lucarini10,lucariniea22} and whose actual existence in the Neoproterozoic Era was revealed by proxy data \cite[e.g.,][]{hoffman98,perre_ea11,herwartz_ea15}. The typical hysteresis behavior of this bistable system is evident in the diagram. 

\section*{REFERENCES}

%\nocite{*}
%\bibliography{aipsamp}% Produces the bibliography via BibTeX.
\bibliography{DE-SR}% Produces the bibliography via BibTeX.

\end{document}